\newcommand*{\ARXIV}{}

\ifdefined\ARXIV 
    \documentclass{agujournal2019}
\else
    \documentclass[draft]{agujournal2019}
\fi

\usepackage[utf8]{inputenc}
\usepackage{url} %this package should fix any errors with URLs in refs.
\usepackage{lineno}
\usepackage{soul}
\usepackage{mathtools}

\ifdefined\ARXIV
\else
\linenumbers
\fi

\draftfalse

\journalname{Geophysical Research Letters}

\begin{document}

\hyphenation{EUCLID}

\title{Thunderstorm nowcasting with deep learning:\\a multi-hazard data fusion model}
\authors{
    Jussi Leinonen\affil{1},
    Ulrich Hamann\affil{1},
    Ioannis V. Sideris\affil{1}~and
    Urs Germann\affil{1} 
}
\affiliation{1}{Federal Office of Meteorology and Climatology MeteoSwiss, Locarno-Monti, Switzerland}

\correspondingauthor{Jussi Leinonen}{jussi.leinonen@meteoswiss.ch}

\begin{keypoints}
    \item We present a deep learning model for nowcasting thunderstorm hazards, and demonstrate it for lightning, hail and heavy precipitation.
    \item The model can provide probabilistic warnings of these hazards on a two-dimensional grid.
    \item We analyze the importance of the different data sources used in the model using explainable artificial intelligence methods.
\end{keypoints}

\begin{abstract}
Predictions of thunderstorm-related hazards are needed in several sectors, including first responders, infrastructure management and aviation. To address this need, we present a deep learning model that can be adapted to different hazard types. The model can utilize multiple data sources; we use data from weather radar, lightning detection, satellite visible/infrared imagery, numerical weather prediction and digital elevation models. We demonstrate the ability of the model to predict lightning, hail and heavy precipitation probabilistically on a 1 km resolution grid, with a temporal resolution of 5 min and lead times up to 60 min. Shapley values quantify the importance of the different data sources, showing that the weather radar products are the most important predictors for all three hazard types.
\end{abstract}

\section*{Plain Language Summary}
Thunderstorms are hazardous to both people and property through various extreme weather phenomena. Predicting these hazards allows individual people, infrastructure managers and emergency services to take action in advance. To serve their needs, we use a model based on an artificial intelligence (AI) to predict the probability of the hazards occurring at a given time and place during the next 60 minutes. The model uses multiple sources of weather observations and predictions to construct the predictions, and can be adapted to function also when some of these sources are unavailable, increasing its reliability. We show that the model can predict the occurrence of lightning, hail and heavy precipitation, detecting and predicting the motion of thunderstorms as well as whether they are increasing or decreasing in severity. We use explainable AI methods to determine how much each of the data sources contributes to the predictions, showing that weather radar observations are the most important source of predictors.

\section{Introduction}

Thunderstorms frequently endanger lives and cause damage to property through several distinct physical processes. Lightning are directly hazardous to humans, igniting fires and disrupting electric infrastructure and aviation \cite{Holle2014Lightning,Holle2016Lightning}. The heavy rain frequently produced by these storms can cause flash floods and trigger landslides \cite{Smith1996RapidanStorm,Davis2001FlashFlood}. Thunderstorms also often produce hail, which can damage buildings, vehicles, crops and forests and is particularly dangerous to flying aircraft \cite{Hohl2002Hailfall,Battaglia2019HailDamage}. They are furthermore frequently associated with strong, gusty wind \cite{Kelly1984NontornadicThunderstorm,Dotzek2003,Allen2016ThunderstormsAustralia}. Thunderstorms are relatively compact objects that develop rapidly, and consequently, the effects of these hazards are sudden and highly localized. Timely, accurate and actionable short-term predictions of them are needed to issue early warnings to the general population, emergency services and infrastructure operators.

Numerical weather prediction (NWP) can forecast thunderstorm occurrence and their typical intensity level in a wider area, but have difficulty predicting the exact time and location where thunderstorms will appear \cite{Sun2014NWPPrecip}. On the very short term of minutes to a few hours, one can instead use \emph{nowcasting}: statistical prediction using the latest observational data. In contrast to NWP models that take hours to run, nowcasting models are typically designed to produce a prediction in seconds, making them suitable for time-critical short-term warning systems. The advantages of the two approaches can also be combined in a \emph{seamless} forecast. As nowcasting systems tend to be accurate on the short term but lose predictive power faster than NWP models, seamless forecasts typically use nowcasting on the short term, NWP forecasts on the long term, and a combination of the two in between \cite{Nerini2019EnsembleNowcast,Nerini2019KalmanNowcasts}.

In the recent years, deep learning (DL) has been increasingly adopted in thunderstorm nowcasting algorithms \cite{Zhou2020LightningDL,Geng2021LightningForecast,Pan2021DLConvective,Cuomo2022StormNowcasting}. \citeA{Leinonen2022Lightning} (henceforth LHG2022) introduced a method based on recurrent-convolutional DL to produce grid-based nowcasts of lightning occurrence at $5\,\mathrm{min}$ temporal resolution over $60\,\mathrm{min}$ using data from multiple sources. The developed network architecture was generalizable and had potential for adaptation to nowcast other hazards. Here, we generalize the architecture and methodology to nowcast different thunderstorm hazards  with the same network architecture. We demonstrate the following capabilities:
\begin{enumerate}
    \item Predicting multiple hazards --- namely, hail and heavy precipitation in addition to lightning --- using the same network structure trained with different target variables.
    \item Training the network with a probabilistic rather than deterministic target.
    \item Providing warnings for multiple threshold levels at once.
    \item Producing warnings based on accumulation over a period longer than the native $5\,\mathrm{min}$ temporal resolution of the DL model.
\end{enumerate}
Similarly to \citeA{Leinonen2021DataSources}, we also investigate how the different data sources contribute to predicting the various hazards and how these contributions vary with increasing nowcast lead time.

\section{Data} \label{sect:data}

To train the DL model, we used the dataset of \citeA{Leinonen2022Weights}. Below, we summarize the contents and processing of the dataset. For further details, we direct the reader to the description in LHG2022.

\subsection{Data sources and processing} \label{sect:data-sources}

We trained the model to combine information from five different sources of data:
\begin{description}
\item[Weather radar] observations were collected from the Swiss operational network \cite{Germann2016SwissRadar,Germann2022RadarOrography}. These data include radar-measured information about the precipitation rate and the vertical structure of the radar reflectivity, such as echo top heights and the vertically integrated liquid water content, at $1\ \mathrm{km}$ horizontal resolution.
\item[Geostationary satellite] imagery was obtained from the Spinning Enhanced Visible and InfraRed Imager \cite<SEVIRI;>{Schmid2000SEVIRI} on the MeteoSat Second Generation 3 (MSG-3) satellite. We used the radiances and brightness temperatures from the visible and infrared (IR) bands; the native resolution of these in the study area is approximately $1\ \mathrm{km} \times 2\ \mathrm{km}$ for the high-resolution visible (HRV) band and $3\ \mathrm{km} \times 5\ \mathrm{km}$ for the others. The bands that consist mostly of reflected solar radiation were normalized with the function $f(x) = x / \cos \theta$, where $\theta$ is the solar zenith angle; these bands are unavailable at night. Furthermore, we used the Nowcasting Satellite Application Facility (NWCSAF) cloud top height, cloud top temperature, cloud optical thickness and cloud top phase products \cite{Derrien2005SeviriCloudMask,Hamann2014SEVIRICloudTop,LeGleau2016CloudATB}.
\item[Lightning detection] measurements were collected by the European Cooperation for Lightning Detection (EUCLID) network of lightning antennas \cite{Schulz2016EUCLID,Poelman2016EUCLID} and delivered by Météorage. The original data consist of locations and various properties of lightning strikes. We aggregated these into maps of lightning density and current density, as well as binary occurrence maps used in lightning prediction. 
\item[NWP] forecasts originated from the Consortium for Small Scale Modelling (COSMO) model \cite{Baldauf2011COSMO} used operationally at MeteoSwiss. We selected various COSMO outputs relevant to thunderstorms, such as the convective available potential energy (CAPE).
\item[Digital elevation model] (DEM) data were from the Advanced Spaceborne Thermal Emission and Reflection Radiometer (ASTER) global DEM \cite{Abrams2020ASTERV3} used to model topography in COSMO.
\end{description}

\subsection{Spatiotemporal characteristics}

The data were collected from an area centered on Switzerland and containing all Swiss territory as well as surrounding regions, $710\ \mathrm{km}$ in the east--west direction and $640\ \mathrm{km}$ north--south (see Supporting Fig.~S1). The study area is based on the range of the Swiss operational radar network. The region is characterized by highly variable topography, ranging from flat plains to mountains over $4500\ \mathrm{m}$ high. The activity of thunderstorms in the region is among the highest in Europe and they regularly cause large property losses, injuries and even deaths \cite{Hilker2009FloodLandslide,Ozturk2018FlashFloods,Taszarek2019ThunderstormClimatology,Feldmann2021Mesocyclone,Kopp2022Hailstorms}.

The data collection period ranged from April to September of 2020. In order to avoid training the model with too many nonconvective cases, we downselected the data to prioritize locations and times near convective activity: For every region where the radar-derived rain rate exceeded $10\ \mathrm{mm\,h^{-1}}$, we included a box of $256\ \mathrm{km} \times 256\ \mathrm{km}$ spatially and $\pm 2\ \mathrm{h}$ temporally in the dataset. Approximately $10^6$ different training samples can be generated from the data, although there is considerable overlap between these; see LHG2022 for more details on the sampling. $10\%$ of the data were used for validation, another $10\%$ for testing, and the rest for training; to minimize correlation between the subsets, each day in the time period was assigned entirely to only one of them.

The data were processed to $5\ \mathrm{min}$ temporal resolution, the native resolution of the radar and satellite datasets. The lightning data were aggregated to $5\ \mathrm{min}$ resolution. The NWP forecasts are provided at a temporal resolution of $1\ \mathrm{h}$ and were interpolated linearly to $5\ \mathrm{min}$ steps. We used $1\ \mathrm{km}$ spatial resolution for the radar, lightning, DEM and HRV, and $4\ \mathrm{km}$ for the remaining variables.

\subsection{Target variables} \label{sect:targets}

In this study, we generalized the DL model of LHG2022 to predict hail and heavy precipitation in addition to lightning. In order to keep our findings compatible with LHG2022, we define lightning occurrence identically: as a binary variable that is set to $1$ if lightning occurred within $8\ \mathrm{km}$ of the grid point within the last $10\ \mathrm{min}$, and $0$ otherwise.

Direct measurements of hail are only available from individual stations, but occurrence of hail is well indicated by the difference of the radar $45\ \mathrm{dBZ}$ echo top height and the freezing level height \cite{Waldvogel1979POH,Foote2005Hail,Barras2019Hail}. For hail, our target variable is the Probability of Hail (POH) product from the weather radar data. This uses the formula of \citeA{Foote2005Hail} for the probability of hail occurrence. We preserve the probabilistic nature of the POH and define our target as a continuous value in the range $[0,1]$.

For heavy precipitation, we derive the target from CombiPrecip \cite{Sideris2014CombiPrecip,Sideris2014CombiPrecipExperience}, which combines radar and raingauge observations to obtain a more accurate and unbiased estimate of precipitation than that of radar alone. As our motivation is to develop a model for issuing short-term warnings of extreme weather, we treat precipitation prediction as a multiclass classification problem, where the model outputs the probabilities of $R$ being between pre-defined threshold levels. We chose thresholds of $R_0=0$, $R_1=10\ \mathrm{mm}$, $R_2=30\ \mathrm{mm}$ and $R_3=50\ \mathrm{mm}$ aggregated over $60\ \mathrm{min}$ at each $1\ \mathrm{km^2}$ grid point, motivated by the warning levels used by MeteoSwiss forecasters. Recognizing that there is inherent uncertainty in the CombiPrecip estimate, we treat it probabilistically and assign probabilities $q_c$ to each class $c \in [0,3]$ as
\begin{linenomath*}
\begin{equation}
q_c = \int_{R_c}^{R_{c+1}} p(R)\,\mathrm{d}R \label{eq:prob_precip}
\end{equation}
\end{linenomath*}
(with $R_4 \coloneqq \infty$) where $p$ is a lognormal probability distribution function. Since CombiPrecip is designed to eliminate systematic biases in the precipitation measurement, we interpret the CombiPrecip estimate as the expected value $\mathrm{E}[R]$. To derive the variance, we analyzed a collection of raingauge measurements collocated with CombiPrecip such that the raingauge used for comparison was omitted from the CombiPrecip analysis. Using the method of \citeA{Ciach1999RadarError} to separate the error due to the lack of raingauge representativeness from that due to uncertainty in the radar measurement, we approximate the standard deviation as
\begin{linenomath*}
\begin{equation}
\mathrm{Std}[R]=0.33\,\mathrm{E}[R].
\end{equation}
\end{linenomath*}
Given $\mathrm{E}[R]$ and $\mathrm{Std}[R]$, the parameters of $p$ can be computed. We used this probabilistic treatment of precipitation to improve the ability of the model to capture extreme events since, for example, a value of $\mathrm{E}[R]=40\ \mathrm{mm}$ gives a nonzero probability also of precipitation exceeding $50\ \mathrm{mm}$. Since most damages caused by rain are due to accumulation of rainwater over times longer than our resolution of $5\ \mathrm{min}$, we predict the $1$-h accumulated precipitation rather than that at each time step.

\section{Methods} \label{sect:models}

\subsection{Neural network} \label{sect:network}

LHG2022 described a recurrent-convolutional DL model for predicting lightning occurrence (see Supporting Fig. S2), based on the model \cite{Leinonen2021Weather4castBigData,Leinonen2021Weather4castStage1} used in the Weather4cast 2021 competition \cite{Herruzo2021Weather4cast} where it outperformed competing architectures such as U-Nets and transformer architectures. We adopt this architecture for each hazard, inheriting the best-performing hyperparameters from LHG2022. The model is built slightly differently for each combination of input data sources, such that only the parts of the model necessary for those inputs are included.

The main architectural change to the DL model in this study is that the prediction of heavy precipitation only uses the last time step of the final layer, which is trained to predict the entire $1$-h accumulation. Furthermore, in contrast to LHG2022 we did not perform model ensembling \cite{Ganaie2021Ensemble} due to the required computational cost.

For lightning, we utilize focal loss \cite{Lin2017FocalLoss} with focusing parameter $\gamma=2$ as the training loss function, so that our results are comparable with LHG2022 where this loss was also adopted. The hail and precipitation targets are defined probabilistically and it is not clear how the focal loss generalizes to such cases. Thus, we use cross entropy (CE) loss, which also performed well in LHG2022 and can be straightforwardly defined for probabilistic targets as:
\begin{linenomath*}
\begin{equation}
    \mathrm{CE}(p,q) = -\sum_c q_c \log(p_c)
\end{equation}
\end{linenomath*}
where $p$ is the predicted probability, $q$ is the target probability and the sum is over the possible classes $c$. In the case of hail, there are two classes (hail or no hail), while with precipitation, there are four classes as defined by Eq.~(\ref{eq:prob_precip}).

Training all $96$ combinations of targets and data sources takes approximately one month on eight Nvidia V100 GPUs. For each target and data source combination, we used the same model architecture and hyperparameters. Ideally, these should be tuned separately for each case to optimize performance, but this would require training each model many times, which would be infeasible with the available resources.

\subsection{Analysis}

We assign a value for the importance of each data source for predicting a hazard, thus making the prediction more explainable, using the Shapley value, introduced by \citeA{Shapley1951} and described in detail by e.g. \citeA[chapter 9.5]{Molnar2022Shapley}. It describes the contribution of a given predictor (or, in our case, a data source) to improving a metric. Given $n_p$ predictors, the Shapley value $\phi_j(M)$ of a predictor $j$ for improving the metric $M$ is 
\begin{equation}
    \phi_j(M) = \sum_{ S \subseteq \{ 1,\ldots,n_p \} \setminus \{j\} } \frac {|S|!\, \left(n_p-|S|-1\right)!}{n_p!} \left(M(S \cup \{j\}) - M(S)\right) \label{eq:shapley}
\end{equation}
where the sum is taken over predictor combinations $S$ that do not contain $j$. Thus, $\phi_j(M)$ is a weighted average of \emph{marginal contributions} 
$M(S \cup \{j\}) - M(S)$, that is, improvements in the metric resulting from adding the predictor to a set of predictors that did not previously contain it. Consequently, $\phi$ has the same units as $M$. The weights are determined using game theory to satisfy multiple properties desirable for a fair interpretation of the value of each predictor:
\begin{enumerate}
    \item The sum of the Shapley values of all predictors $\sum_{j=1}^{n_p} \phi_j(M)$ is equal to the improvement in the metric given by the complete set of predictors.
    \item Two predictors that contribute equally have equal Shapley values.
    \item A predictor that does not change the metric has a Shapley value of $0$.
    \item The Shapley value of a predictor for an average of multiple models is the average of the Shapley values of that predictor for the individual models.
\end{enumerate}
Determining the Shapley values using Eq.~\ref{eq:shapley} requires the computation of $M(S)$ for each subset $S \subseteq \{ 1,\ldots,n_p \}$, including the empty set. Since we consider the five data sources described in Sect.~\ref{sect:data-sources}, we need to train the model $2^5=32$ times, and therefore can compute the Shapley values directly without resorting to approximations, which become necessary for larger numbers of predictors.

\section{Results and discussion} \label{sect:results}

\subsection{Prediction}

\begin{figure*}
    \noindent\includegraphics[width=0.95\textwidth]{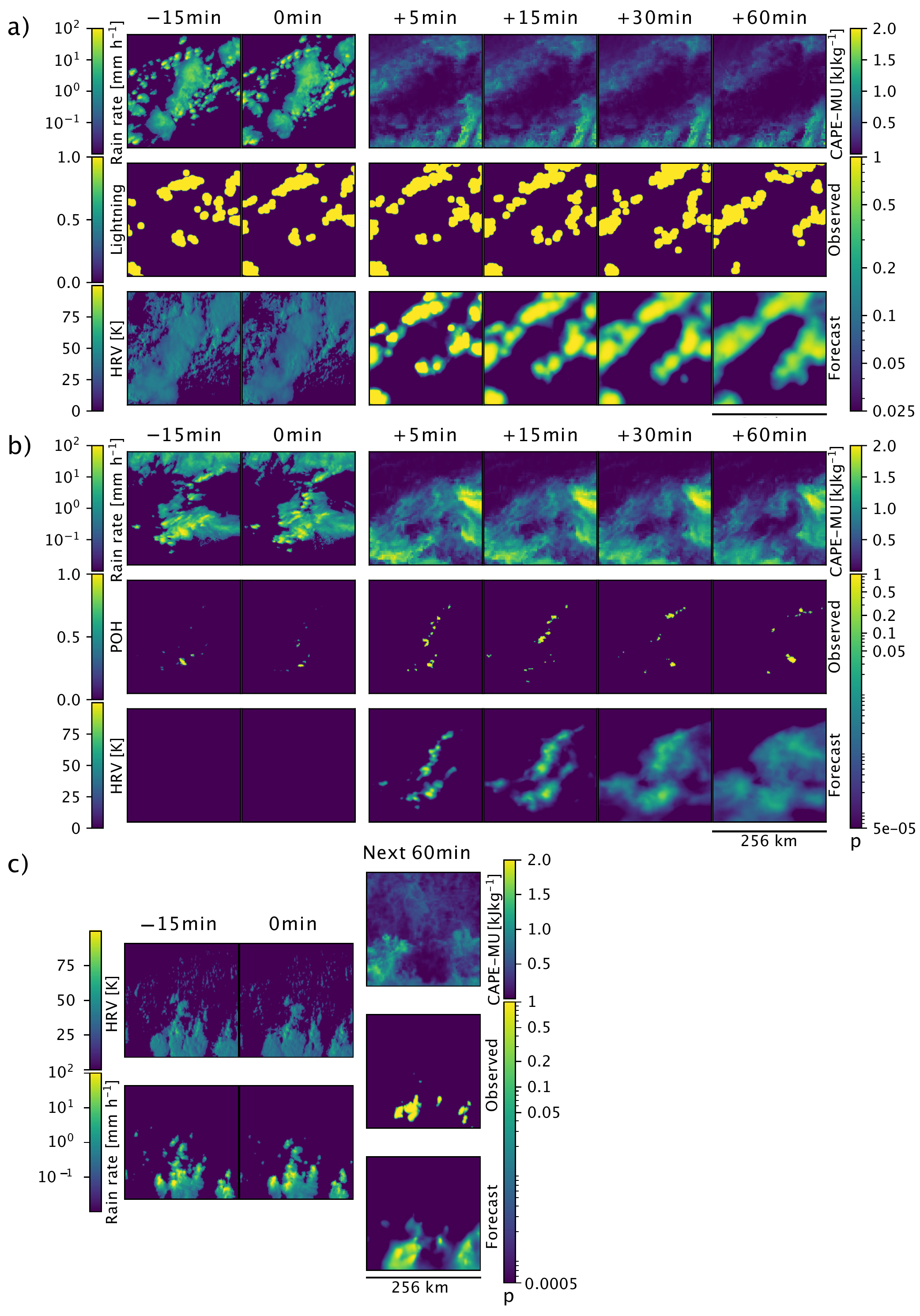}
    \caption{a) The results of lightning prediction. On the left, three input variables (rain rate, lightning, and the satellite HRV channel) are shown at two time steps. On the right, future time steps are shown, with the first row giving the CAPE, the second showing the observed lightning, and the third showing the model prediction of lightning occurrence probability.\\
    b) As (a), but showing the observed and predicted POH instead of lightning. The HRV is not shown as the case occurs at night.\\
    c) As (a) and~(b), but showing the probability of accumulated precipitation in the next 60 minutes exceeding $10\ \mathrm{mm}$. Consequently, only one future time step is shown.}
    \label{fig:sample-all}
\end{figure*}

Examples of the prediction of lightning, hail and heavy precipitation are shown in Fig.~\ref{fig:sample-all}. The example of lightning prediction in Fig.~\ref{fig:sample-all}a shows a case where the location of lightning activity in relatively slowly moving thunderstorms is forecast by the model for the 60-min period. A more extensive discussion of similar cases can be found in LHG2022.

An example of the prediction of hail is shown in Fig.~\ref{fig:sample-all}b. The model predicts the imminent occurrence of hail on the top half of the image even though the past time steps contain very few pixels with nonzero POH on the top half. Compared to the lightning prediction, the shorter time scale of the predictability of hail is apparent in this example, as the predicted location of the hail becomes very diffuse at the $+60\ \mathrm{min}$ time step. Nevertheless, the model appears able to distinguish wider areas with a general risk of hail from those where hail is very unlikely (e.g. on the top left).

Figure~\ref{fig:sample-all}c shows a case of the prediction of heavy precipitation. Since we predict the total accumulation of precipitation during the next $60\ \mathrm{min}$, we only show a single prediction instead of multiple time steps. As described in Sect.~\ref{sect:targets}, the probability of precipitation is predicted in four classes. For the purposes of visualization, we sum the probabilities of the last three classes to visualize the probability of $R>10\ \mathrm{mm}$. Also in this case, we see that the model predicts the location of the heaviest precipitation by assigning high probabilities of exceeding $10\ \mathrm{mm}$ to those areas where the observation shows that this is actually the case.

Over the entire test dataset, we achieve the following pixelwise critical success indices (CSI) when the warning threshold is selected to give optimal CSI:
\begin{itemize}
    \item Lightning: $0.767$ ($5\ \mathrm{min}$), $0.544$ ($15\ \mathrm{min}$), $0.443$ ($30\ \mathrm{min}$), $0.317$ ($60\ \mathrm{min}$)
    \item Hail: $0.454$ ($5\ \mathrm{min}$), $0.267$ ($15\ \mathrm{min}$), $0.140$ ($30\ \mathrm{min}$), $0.057$ ($60\ \mathrm{min}$)
    \item Rain ($60\ \mathrm{min}$ accumulation): $0.320$ ($R>10\ \mathrm{mm}$), $0.182$ ($R>30\ \mathrm{mm}$), $0.131$ ($R>50\ \mathrm{mm}$)
\end{itemize}
In agreement with the example cases, hail and precipitation are more challenging to predict than lightning. This is explained by time scales of the different variables: The correlation time scales of the variables are $27.1\ \mathrm{min}$ for lightning, $8.8\ \mathrm{min}$ for hail and $13.7\ \mathrm{min}$ for $R>10\ \mathrm{mm\,h^{-1}}$. The longer time scale of lightning is at least partially due to the smoothing caused by the ``within $8\ \mathrm{km}$ in the last $10\ \mathrm{min}$'' definition, in contrast to the precipitation and hail that are predicted pixelwise. We have included further metrics and an autocorrelation plot in the Supporting Information to support this analysis. The confusion matrices for all models are available in \citeA{Leinonen2022Pretrained}.

\subsection{Data source importance} \label{sect:importance}

\begin{figure*}
    \noindent\includegraphics[width=\textwidth]{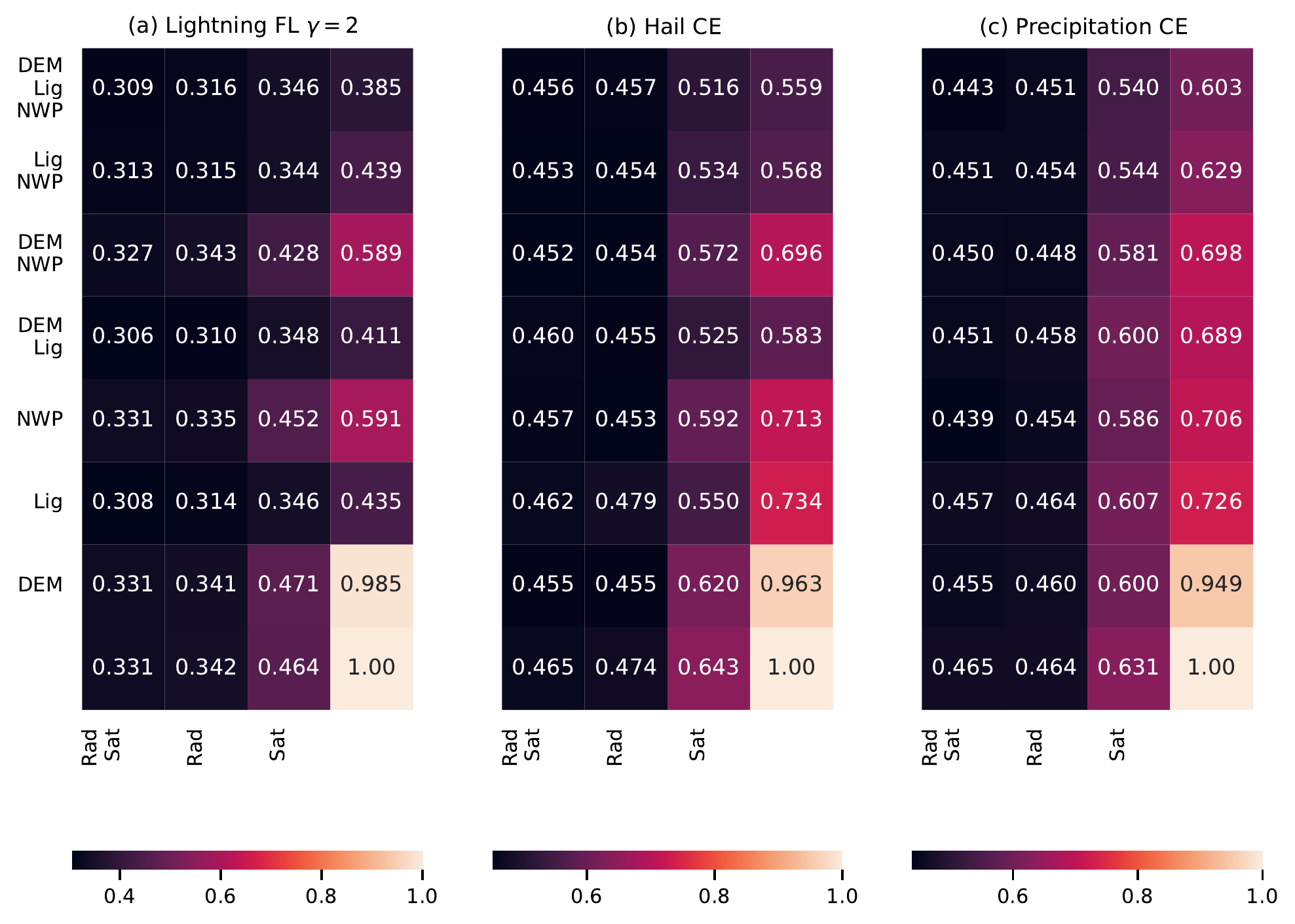}
    \caption{The average loss in the test dataset for the prediction of (a) lightning, (b) hail and (c) heavy precipitation using different data sources. In every panel, each square shows the loss for the data sources indicated by the combination of the labels for the corresponding row and column (abbreviated to ``Rad'' for radar, ``Sat'' for satellite and ``Lig'' for lightning). For example, the top left corner of each panel shows the model trained with all five data sources, the bottom left shows the model using only radar and satellite data, and the bottom right shows the ``ignorant'' model that did not receive any inputs. All loss scores are scaled such that the loss of the ignorant model is set to $1$.}
    \label{fig:metrics-sources}
\end{figure*}

Figure~\ref{fig:metrics-sources} shows the loss over the test dataset for lightning, hail and precipitation prediction. In the losses for lightning (Fig.~\ref{fig:metrics-sources}a), one can see the importance of the radar observations from the leftmost two columns, and the importance of the lightning observations from the first, second, fourth and sixth rows, which correspond to combinations of data sources where lightning observations are included. Satellite observations are less powerful predictors than those from radar, but can still perform well, especially when supported by lightning observations. The model based only on NWP forecasts has considerable skill over the ignorant model on its own, but it does not seem to add much to the skill of the models that make use of observational data. The benefits from the DEM appear to be minimal at best.

The pattern of losses for hail (Fig.~\ref{fig:metrics-sources}b) is similar to that for lightning, except the radar observations are more dominant. In fact, all models using the radar data perform substantially better than any of the models without them. The lightning observations are less useful for hail than for predicting lightning itself, but still beneficial especially when radar is not available. The satellite observations achieve reasonable skill when complemented by other observations, but do not have a major impact if radar data are already available. Meanwhile, the NWP data yield small but consistent improvements to the loss, as does the DEM, which is somewhat more helpful for predicting hail than lightning.

The results for heavy precipitation are shown in Fig.~\ref{fig:metrics-sources}c. The relative benefit of radar data is somewhat higher than in the case of hail, while that of lightning data is slightly smaller. Otherwise, the patterns are very similar to those for hail.

\subsection{Shapley values}

\begin{figure}
    \noindent\includegraphics[width=\linewidth]{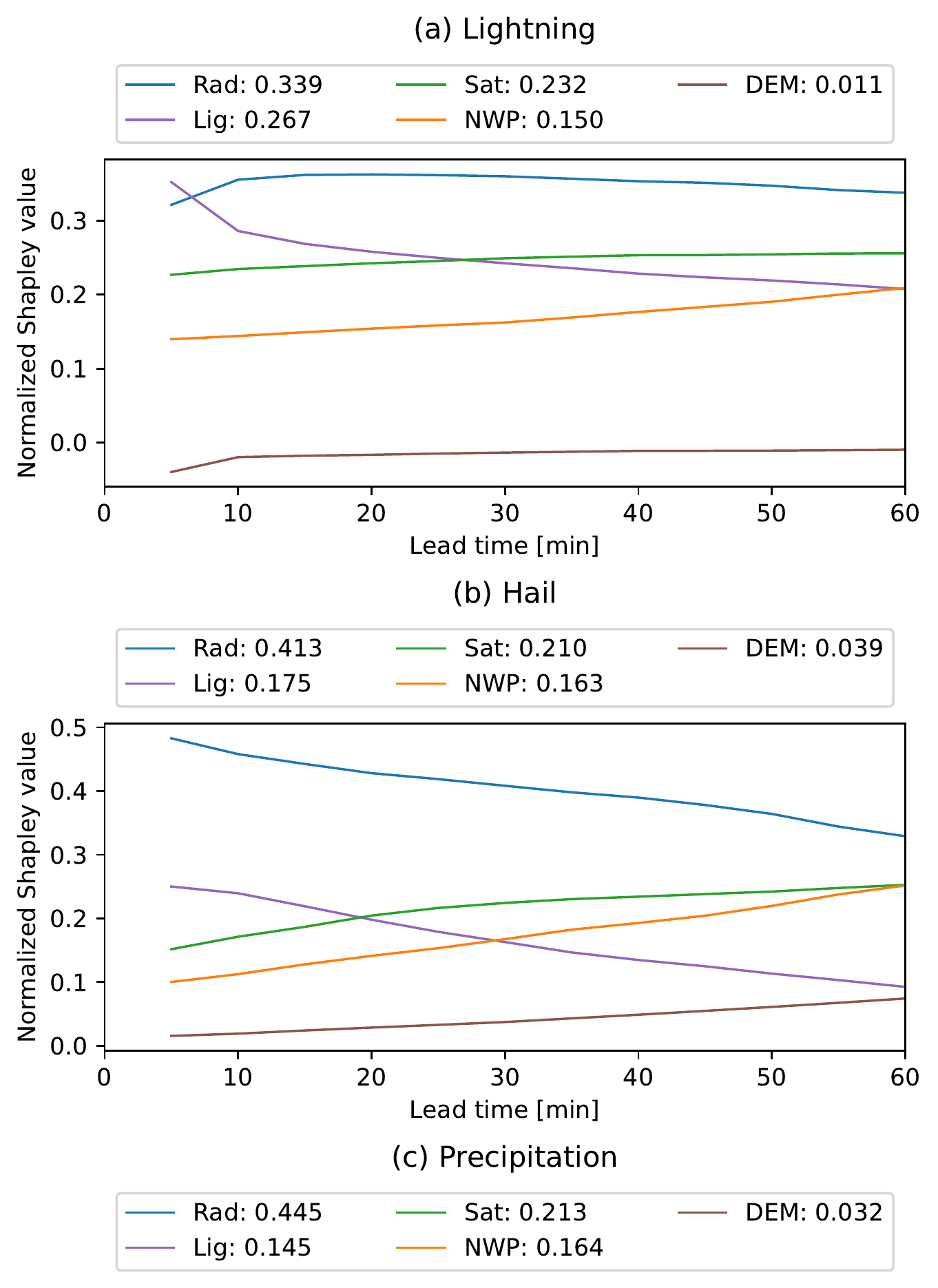}
    \caption{The Shapley values (normalized to a sum of $1$) of the different data sources, evaluated with the test dataset, for the prediction of (a) lightning, (b) hail and (c) heavy precipitation. For lightning and hail, the Shapley values as a function of lead time are also shown; for heavy precipitation, we only predict the $60$-min accumulated precipitation so this cannot be displayed.}
    \label{fig:shapley}
\end{figure}

In Fig.~\ref{fig:shapley}, we show the Shapley values as quantitative indicators of the total importance of each data source, and also plot them against the forecast lead time, allowing us to assess how the importance of each data source changes.

The Shapley values for lightning prediction (Fig.~\ref{fig:shapley}a) agree with our qualitative assessment in Sect.~\ref{sect:importance}. Radar is the most important source, followed by the lightning and satellite data. NWP is less important but nevertheless clearly beneficial. The importance of the DEM is minimal; it is even computed as slightly negative for the first time step, but a closer inspection of the data suggests that this is because a few models using the DEM have coincidentally converged to somewhat worse results due to the stochastic nature of the DL optimization process. The lightning data drop sharply in importance at first; this is because at the first prediction step, the observation directly provides lightning occurrence for the first $5\ \mathrm{min}$ of the $10\ \mathrm{min}$ averaging interval. Then, a more gradual decline in the importance of the lightning data follows. The drop in the lightning importance is compensated at first by increase in the relative importance of the radar, which then turns into a slight decline around $20\ \mathrm{min}$ lead time. The importance of satellite and NWP data instead display a positive trend, eventually passing the lightning data.

For hail (Fig.~\ref{fig:shapley}b), the Shapley values have a similar pattern to those of lightning prediction, with the radar considered even more important than for lightning prediction. Satellite and lightning observations are of similar importance when considered over the whole time period, but they display clearly opposite trends. The importance of the lightning data is second only to radar at short lead times, but declines rapidly; this may be connected to the known ``lightning jump'' phenomenon \cite{Williams1999Lightning,Schultz2009Lightning,Farnell2017Lightning} where a sudden increase in lightning precedes hail by $5$--$30\ \mathrm{min}$. Meanwhile, the relative importance of satellite imagery increases over time. This, along with the similar result for lightning prediction, indicates that satellite visible/IR observations have predictive power over longer timescales than radar observations. NWP data provide a similar benefit as for lightning and, as expected, increase in importance over time. The DEM importance is small but more notable than for lightning.

The importance of the features for heavy precipitation (Fig.~\ref{fig:shapley}c) is again similar to lightning, confirming the conclusions drawn about the results shown in Fig.~\ref{fig:metrics-sources}c.

\section{Discussion and conclusions} \label{sect:conclusions}

In this paper, we have demonstrated a deep learning model architecture that can use a combination of many data sources to predict various hazards from thunderstorms. The model produces probabilistic predictions, allowing end users to choose thresholds for issuing warnings and taking protective action, adapting the threshold to their tolerance for missed events and false alarms. In this work, we considered warnings for lightning, hail and heavy precipitation. The hazard probabilities are produced at a resolution of $5\ \mathrm{min}$ / $1\ \mathrm{km}$. Lightning occurrence is considered as a binary event, while hail prediction is trained with the radar-derived probability of hail as the target variable. Meanwhile, heavy precipitation is predicted using probabilities of four classes and aggregated into 60-min accumulated precipitation. We chose these different approaches for different hazards in order to demonstrate the flexibility of the model, which can be adapted to the users' needs with minor changes to the training procedure. 

Of the three hazard types considered, lightning predictions are the most accurate. Hail predictions have lower accuracy and are not recommended to be used at long lead times at single-pixel scales due to the low prediction accuracy, but they can still indicate the probable occurrence of hail somewhere in a wider area, as demonstrated in Supporting Fig.~S10. Precipitation predictions are qualitatively different from the others because they are given for the 60-min accumulation, but their accuracy can be summarized as being between the hail and lightning predictions.

The data source importance analysis shows that the ground-based radar data are the most important predictor for all three hazards considered, and may be sufficient on their own to make accurate predictions in many applications, especially for heavy precipitation. Lightning detection is important for predicting lightning, but also useful for predicting hail (especially at short lead times) and precipitation. Satellite data are beneficial for all three targets; while they have a lower predictive capability than radar, they retain their predictive power over longer timescales, especially in the case of predicting hail. Geostationary satellite data are also available almost everywhere on Earth, allowing predictions to be made when radar observations are unavailable due to lack of installed radars or temporary outages. Indeed, the good performance of the combination of satellite imagery and lightning data at predicting lightning occurrence (Lig+Sat in Fig.~\ref{fig:metrics-sources}a) suggests that high-performance lightning nowcasting is feasible globally using the latest generation of geostationary weather satellites, e.g. GOES-R series or MeteoSat Third Generation, which have both a visible/IR imager and a lightning mapper. The NWP data are useful for all three targets and, as expected, their relative importance increases at longer lead times. Meanwhile, the DEM data improve the prediction marginally but consistently. The ability of the model to use different combinations of data sources not only improves prediction accuracy, but also increases the resilience of the model in operational use: A model trained with a different combination of data sources can be substituted if some of the data sources are unavailable or delayed.

We recommend later studies to investigate whether the forecast can be improved, particularly at longer lead times, with model architecture improvements but also inputs that better constrain the state of the atmosphere, such as 3D data, polarimetric radar, higher-resolution satellite observations, and data from rapid-update NWP models with complex microphysics. It would be ideal to predict strong winds caused by thunderstorms as well. Unfortunately, we lack a clear ``ground truth'' variable giving the surface wind everywhere on the grid. Wind is measured at weather stations, but these are not very representative especially in thunderstorm conditions. Meanwhile, Doppler radar measurements indicate wind speed, but these are taken hundreds of meters to kilometers above the surface where the biggest impacts occur. Thus, we expect that training DL models for this purpose will require a novel approach.

The lightning predictions are somewhat robust up to at least $60\ \mathrm{min}$, but with the hail predictions, the model loses accuracy more quickly after the first $15\ \mathrm{min}$. Even short warning times still allow many preventive interventions: For example, members of the public who receive warnings through smartphones can seek cover and protect vulnerable property, and automated protection systems such as automatically-raised window blinds can activate. However, even delays of a few minutes may significantly impede the ability to act on such warnings. While the model itself can be executed in a few seconds even without GPU hardware, it is important to host the model and the accompanying data and warning system in a low-latency environment. We expect that in operational use the model will be run by meteorological services who already have such infrastructure in place, and then distributed to the end users. 

\section{Open Research}
The preprocessed training, validation and testing datasets created for this study are available for noncommercial use at \url{https://doi.org/10.5281/zenodo.6802292} \cite{Leinonen2022Weights}. The results, pretrained models and additional data needed to train the hail and lightning models can be found at \url{https://doi.org/10.5281/zenodo.7157986} \cite{Leinonen2022Pretrained}. The DL and analysis code used in this study can be found at \url{https://github.com/MeteoSwiss/c4dl-multi}. The original data from EUCLID lightning network are proprietary and cannot be made available in raw form. The original data from the Swiss radar network and the COSMO NWP model can be made available for research purposes on request. The MSG SEVIRI Rapid Scan radiances are available to EUMETSAT members and participating organizations at the EUMETSAT Data Store (\url{https://data.eumetsat.int/}). The NWCSAF products can be created from these data using the publicly available NWCSAF software available at \url{https://www.nwcsaf.org/}. The ASTER DEM can be obtained from \url{https://doi.org/10.5067/ASTER/ASTGTM.003} \cite{ASTERGDEMV3Data}.

\acknowledgments
JL was supported by the fellowship ``Seamless Artificially Intelligent Thunderstorm Nowcasts'' from the European Organisation for the Exploitation of Meteorological Satellites (EUMETSAT). The hosting institution of this fellowship is MeteoSwiss in Switzerland.

\clearpage

\bibliography{journalabrv,thunderstormrdl}

\end{document}

%% ------------------------------------------------------------------------ %%
%
%  TITLE
%
%% ------------------------------------------------------------------------ %%

%\includegraphics{agu_pubart-white_reduced.eps}

\title{Supporting Information for ``Thunderstorm nowcasting with deep learning: a multi-hazard data fusion model''}
%
% e.g., \title{Supporting Information for "Terrestrial ring current:
% Origin, formation, and decay $\alpha\beta\Gamma\Delta$"}
%
%DOI: 10.1002/%insert paper number here%

%% ------------------------------------------------------------------------ %%
%
%  AUTHORS AND AFFILIATIONS
%
%% ------------------------------------------------------------------------ %%

% List authors by first name or initial followed by last name and
% separated by commas. Use \affil{} to number affiliations, and
% \thanks{} for author notes.
% Additional author notes should be indicated with \thanks{} (for
% example, for current addresses).

% Example: \authors{A. B. Author\affil{1}\thanks{Current address, Antartica}, B. C. Author\affil{2,3}, and D. E.
% Author\affil{3,4}\thanks{Also funded by Monsanto.}}

\authors{
    Jussi Leinonen\affil{1},
    Ulrich Hamann\affil{1},
    Ioannis V. Sideris\affil{1}~and
    Urs Germann\affil{1} 
}
\affiliation{1}{Federal Office of Meteorology and Climatology MeteoSwiss, Locarno-Monti, Switzerland}

%% ------------------------------------------------------------------------ %%
%
%  BEGIN ARTICLE
%
%% ------------------------------------------------------------------------ %%

% The body of the article must start with a \begin{article} command
%
% \end{article} must follow the references section, before the figures
%  and tables.

\begin{article}

%% ------------------------------------------------------------------------ %%
%
%  TEXT
%
%% ------------------------------------------------------------------------ %%

\begin{figure}
    \centerline{\includegraphics[width=0.8\textwidth]{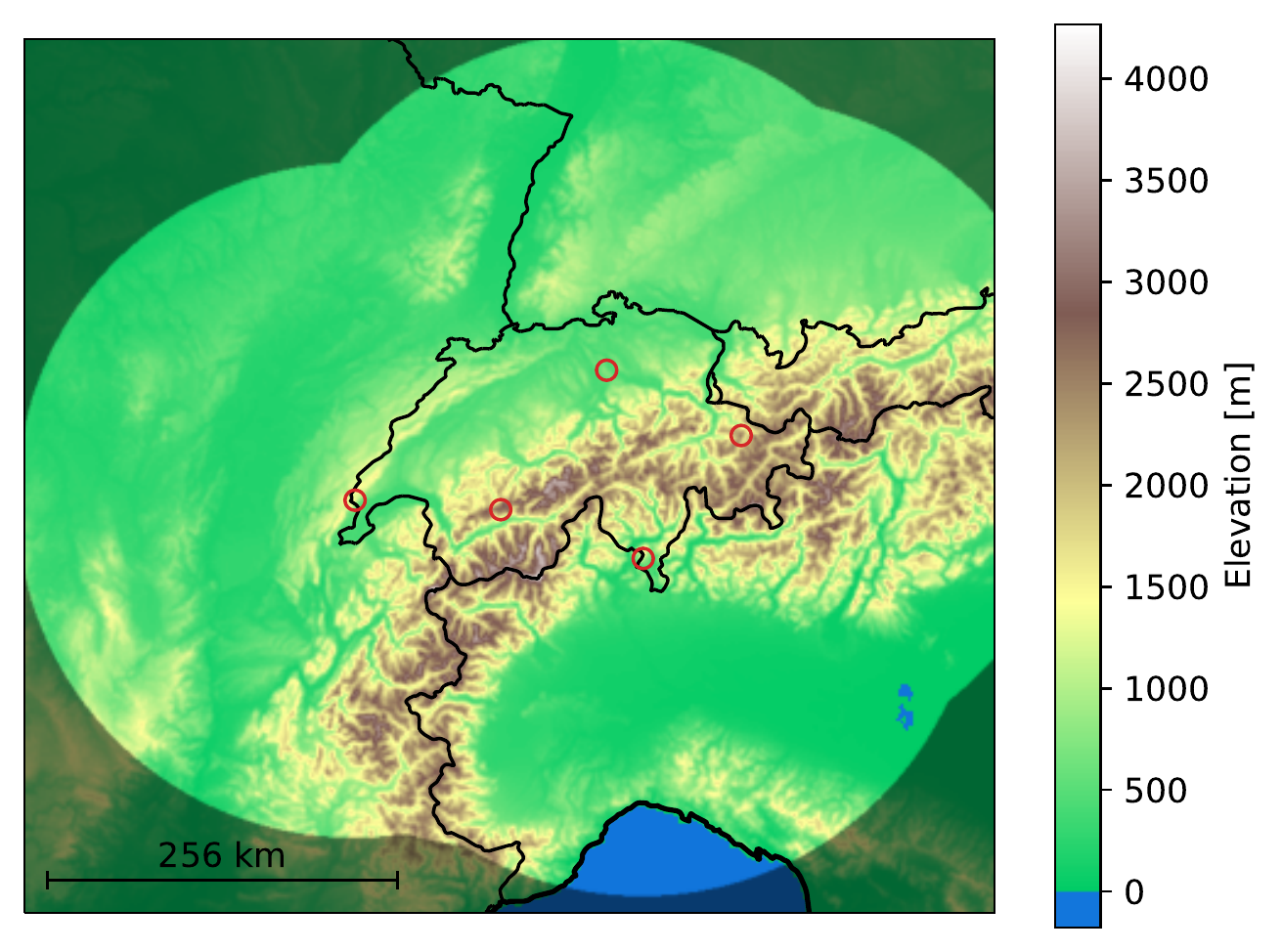}}
    \caption{The study area with the terrain elevation shown in color and the international borders as black lines. The locations of the weather radars are shown as red circles and the shaded area depicts the area outside the range of the radars and hence excluded from the study. The scale bar indicates a distance of 256 km, the size of the subdomains used for training. Reproduced from Leinonen, Hamann and Germann (2022b). \copyright American Meteorological Society. Used with permission.}
    \label{fig:study_area}
\end{figure}

\begin{figure}
    \centerline{\includegraphics[width=\textwidth]{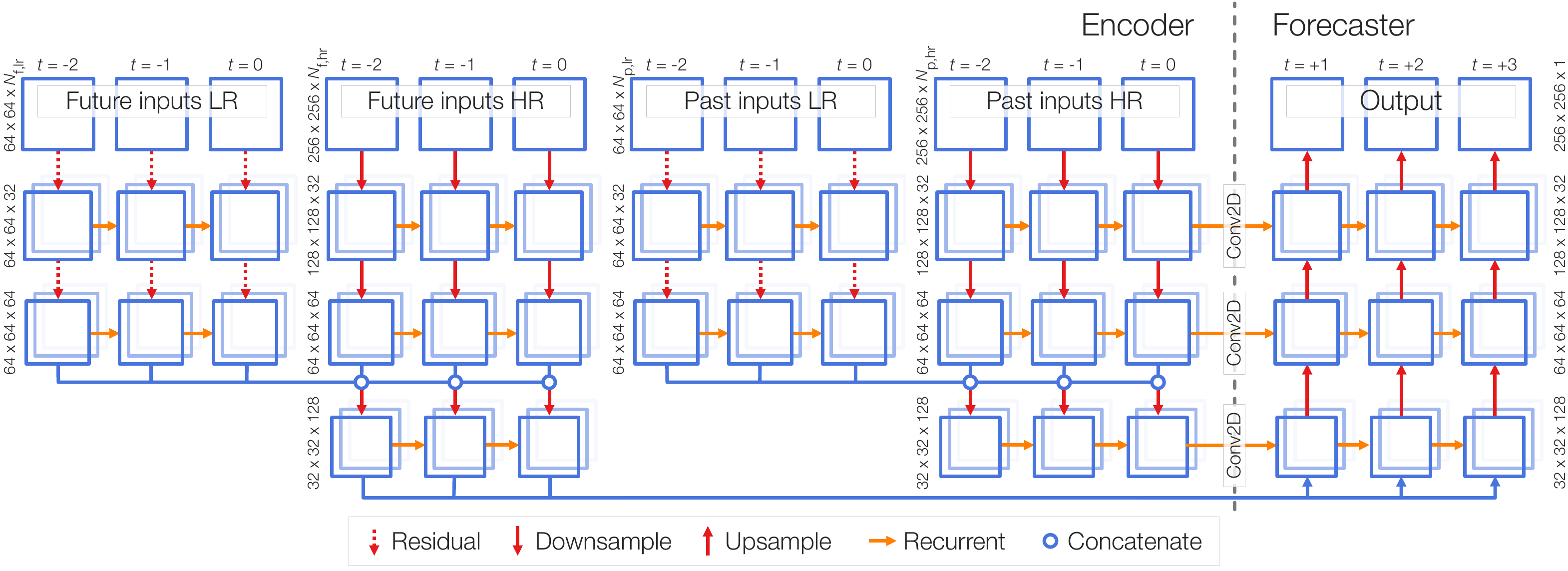}}
    \caption{An illustration of the network architecture. For clarity, only $3$ time steps are shown for both the past and the future; the actual network processes $6$ past time steps and $12$ future time steps. $N$ is the number of predictors; $\mathrm{lr}$ indicates low resolution and $\mathrm{hr}$ high resolution, while $\mathrm{p}$ indicates the past timeframe and $\mathrm{f}$ the future timeframe (i.e. COSMO variables). In our case $N_{\mathrm{f},\mathrm{lr}}=9$, $N_{\mathrm{f},\mathrm{hr}}=10$, $N_{\mathrm{p},\mathrm{lr}}=20$ and $N_{\mathrm{p},\mathrm{hr}}=20$. Reproduced from Leinonen, Hamann and Germann (2022b). \copyright American Meteorological Society. Used with permission.}
    \label{fig:network}
\end{figure}

\begin{figure}
    \centerline{\includegraphics[width=\textwidth]{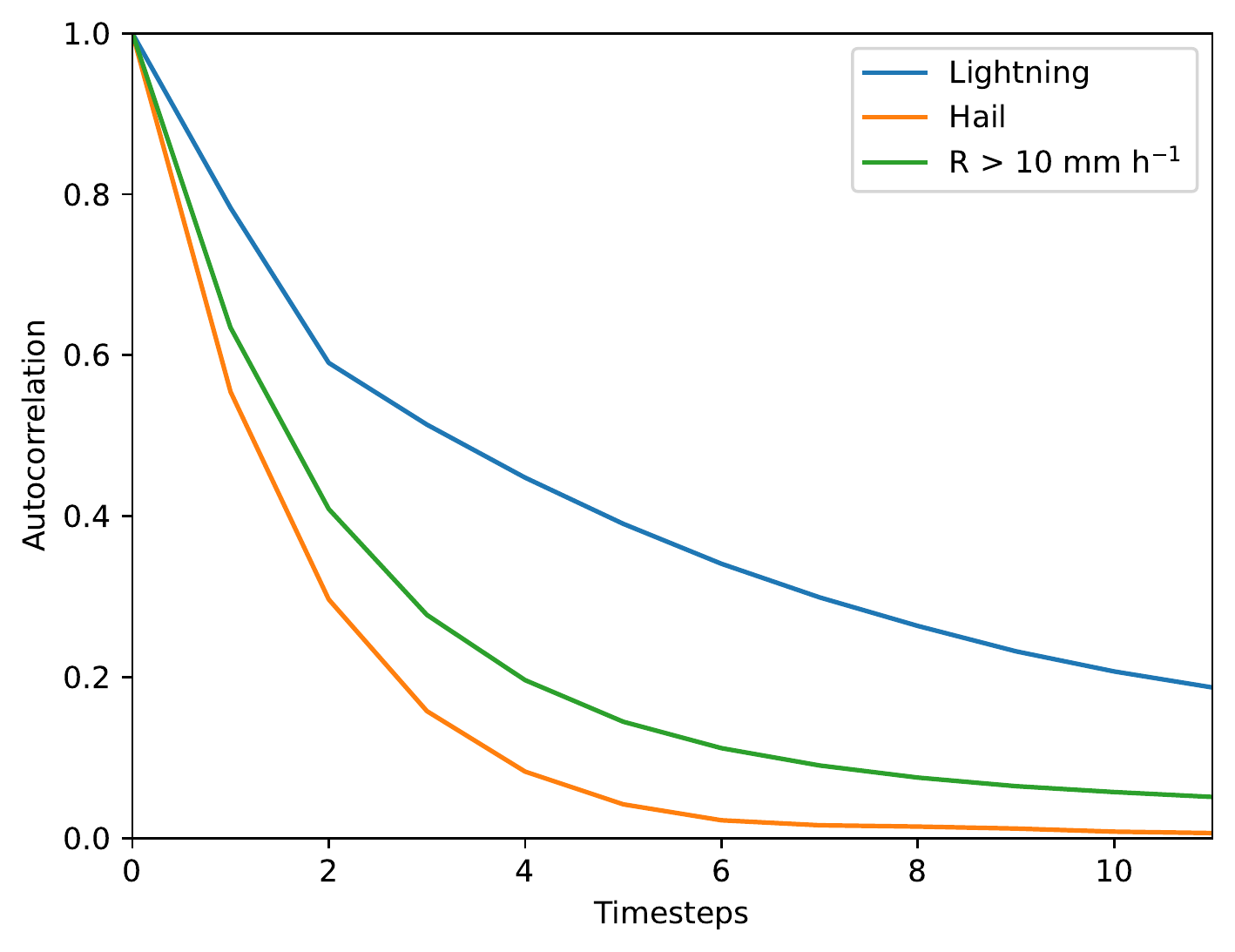}}
    \caption{The temporal autocorrelation of our lightning, hail and precipitation rate $R > 10\ \mathrm{mm\,h^{-1}}$.}
    \label{fig:autocorr}
\end{figure}

\begin{figure}
    \centerline{\includegraphics[width=\textwidth]{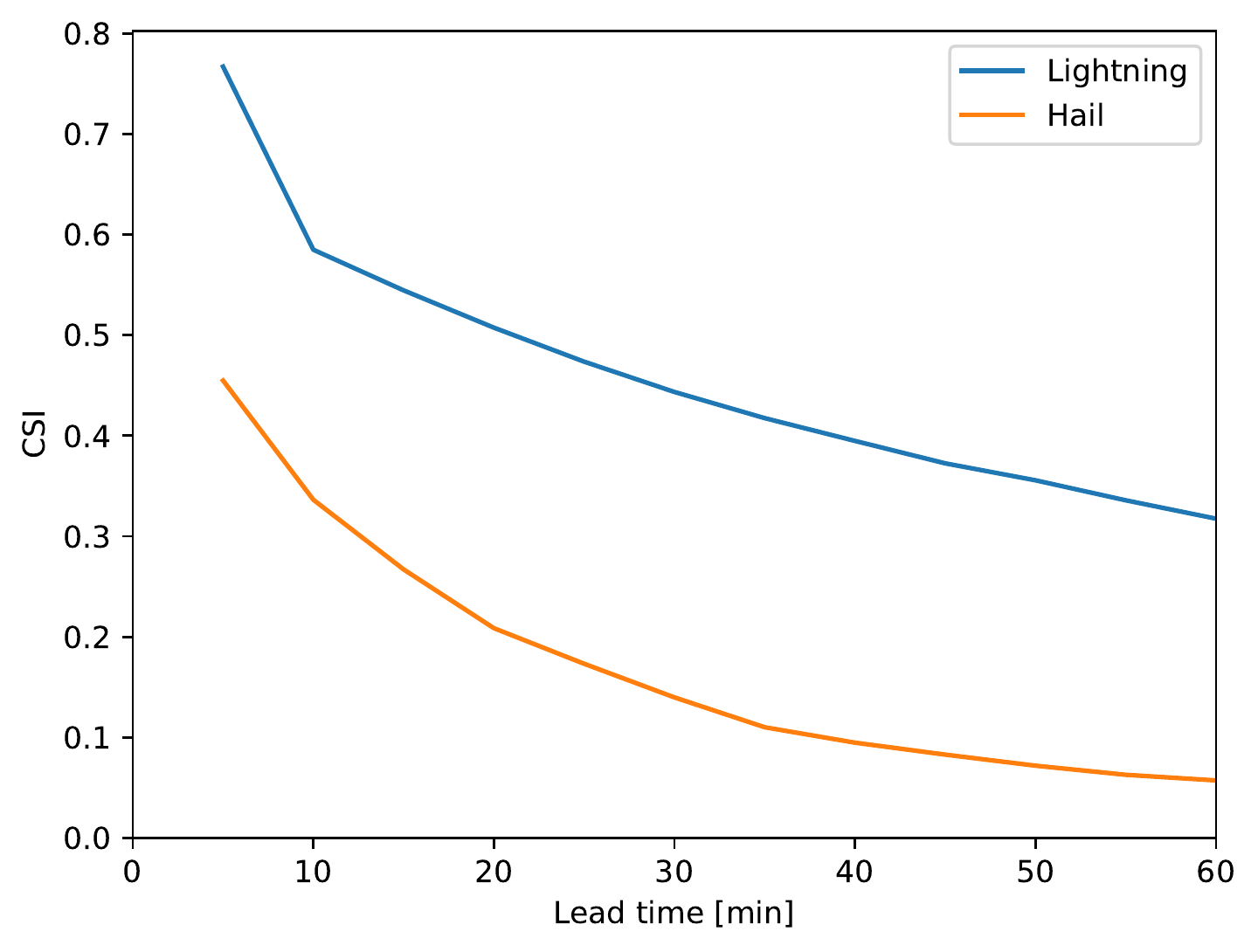}}
    \caption{The critical success index (CSI) of the lightning and hail models as a function of lead time. The probability threshold has been selected to give the optimal metric.}
    \label{fig:CSI-leadtime}
\end{figure}

\begin{figure}
    \centerline{\includegraphics[width=\textwidth]{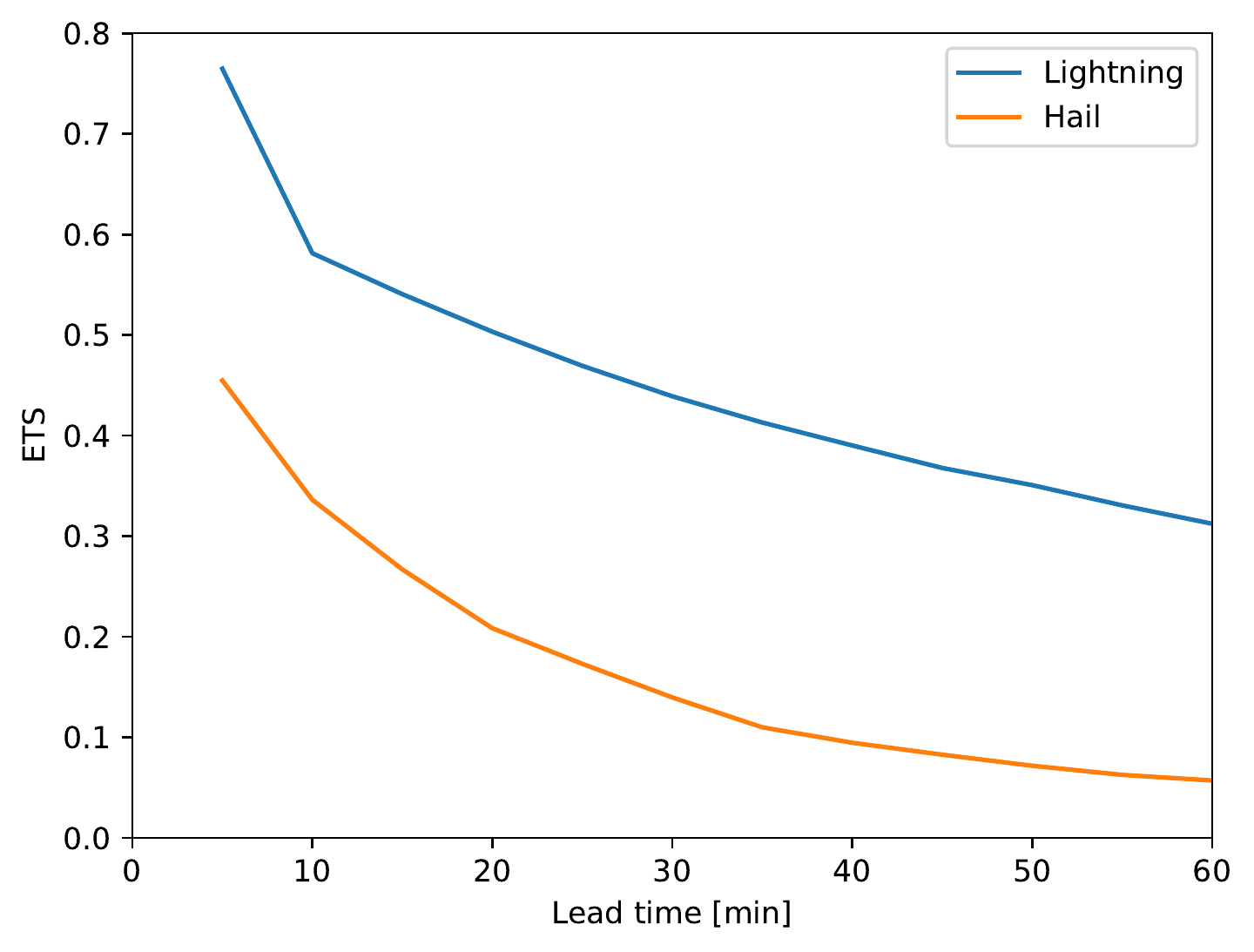}}
    \caption{As Fig.~\ref{fig:CSI-leadtime}, but for the equitable threat score (ETS).}
    \label{fig:ETS-leadtime}
\end{figure}

\begin{figure}
    \centerline{\includegraphics[width=\textwidth]{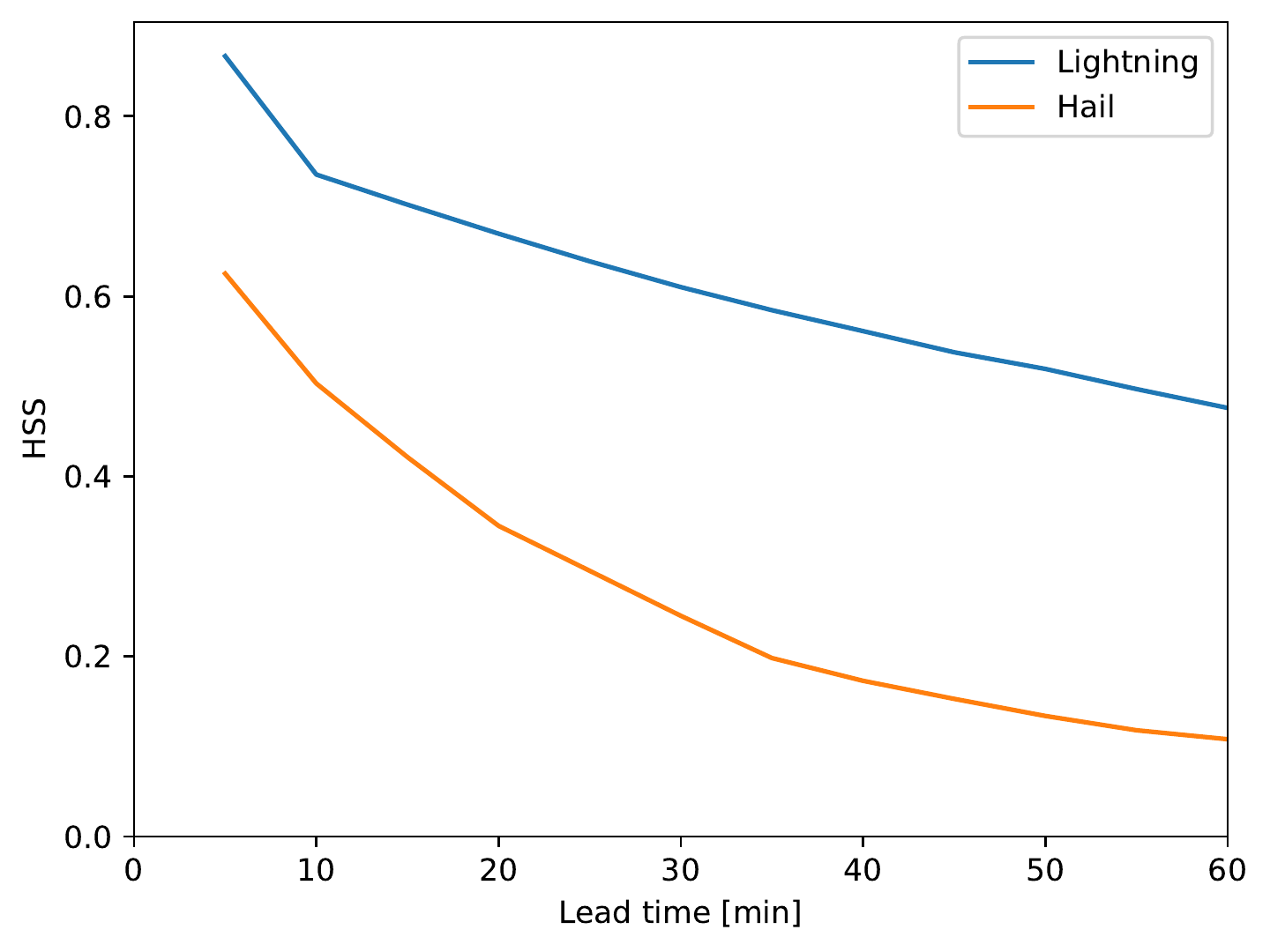}}
    \caption{As Fig.~\ref{fig:CSI-leadtime}, but for the Heidke skill score (HSS).}
    \label{fig:HSS-leadtime}
\end{figure}

\begin{figure}
    \centerline{\includegraphics[width=\textwidth]{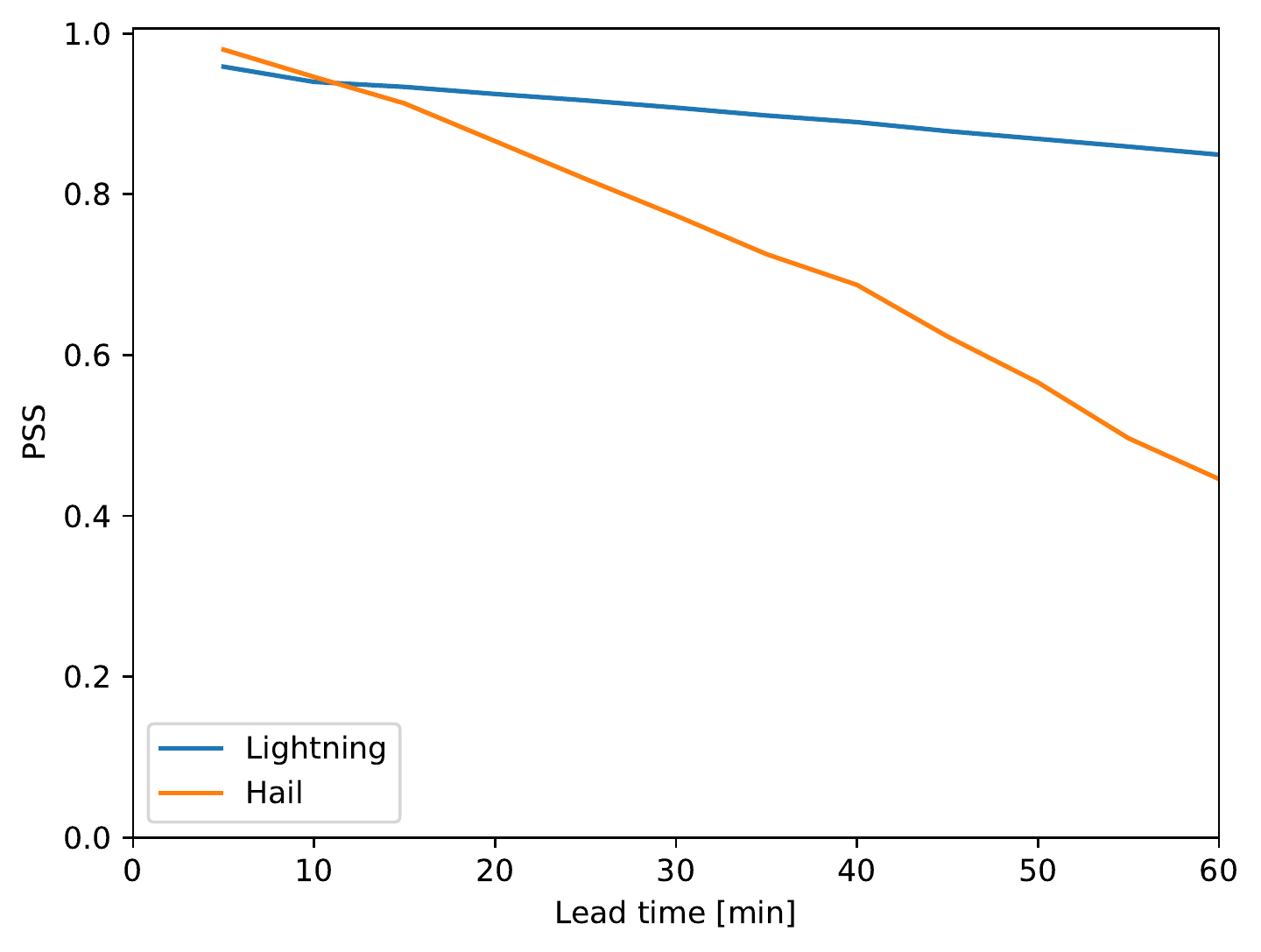}}
    \caption{As Fig.~\ref{fig:CSI-leadtime}, but for the Peirce skill score (PSS).}
    \label{fig:PSS-leadtime}
\end{figure}

\begin{figure}
    \centerline{\includegraphics[width=\textwidth]{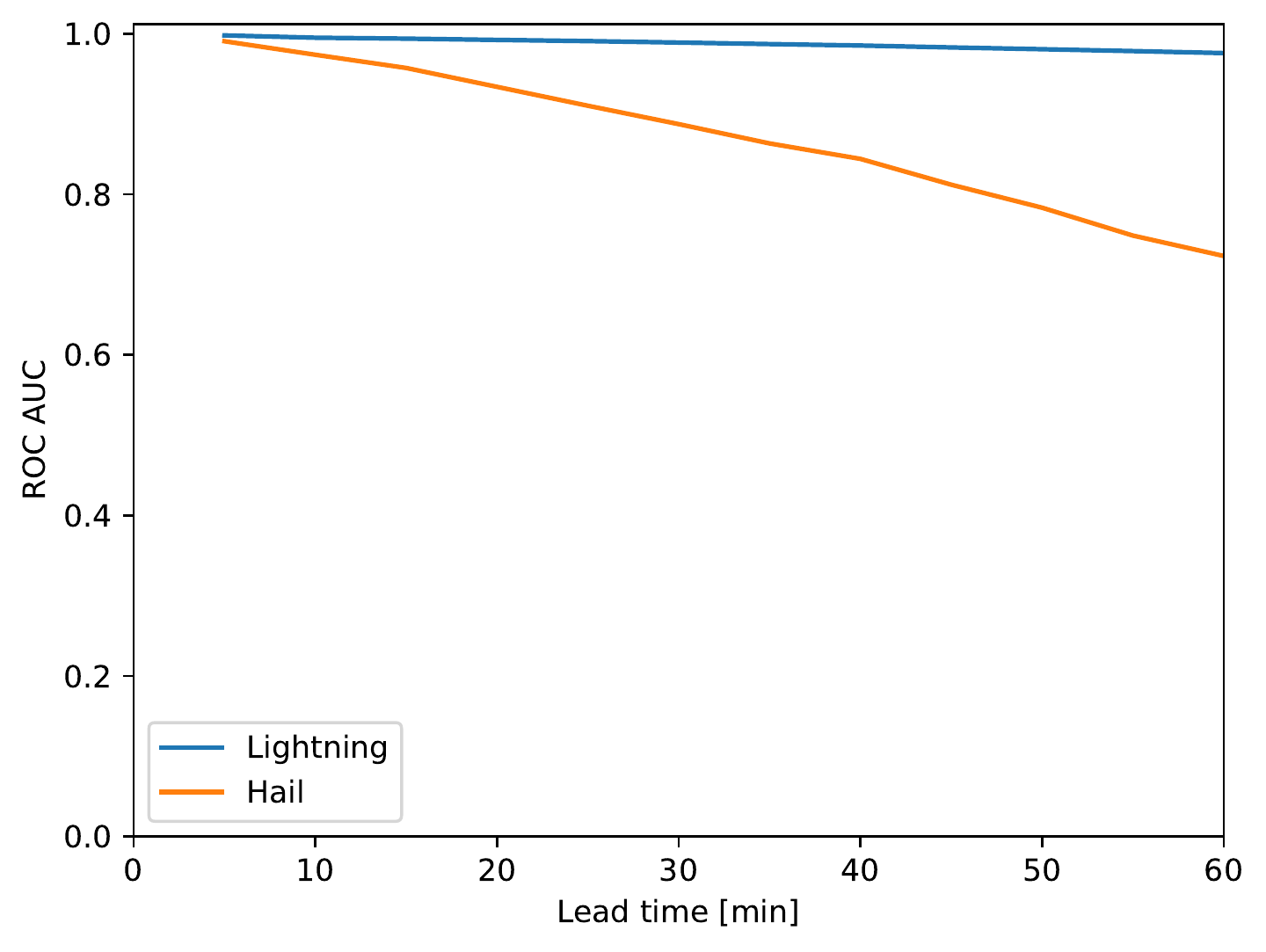}}
    \caption{As Fig.~\ref{fig:CSI-leadtime}, but for the receiver operating characteristic (ROC) area under curve (AUC). The AUC scores are integrated over all thresholds, so selecting a threshold is not necessary.}
    \label{fig:ROC_AUC-leadtime}
\end{figure}

\begin{figure}
    \centerline{\includegraphics[width=\textwidth]{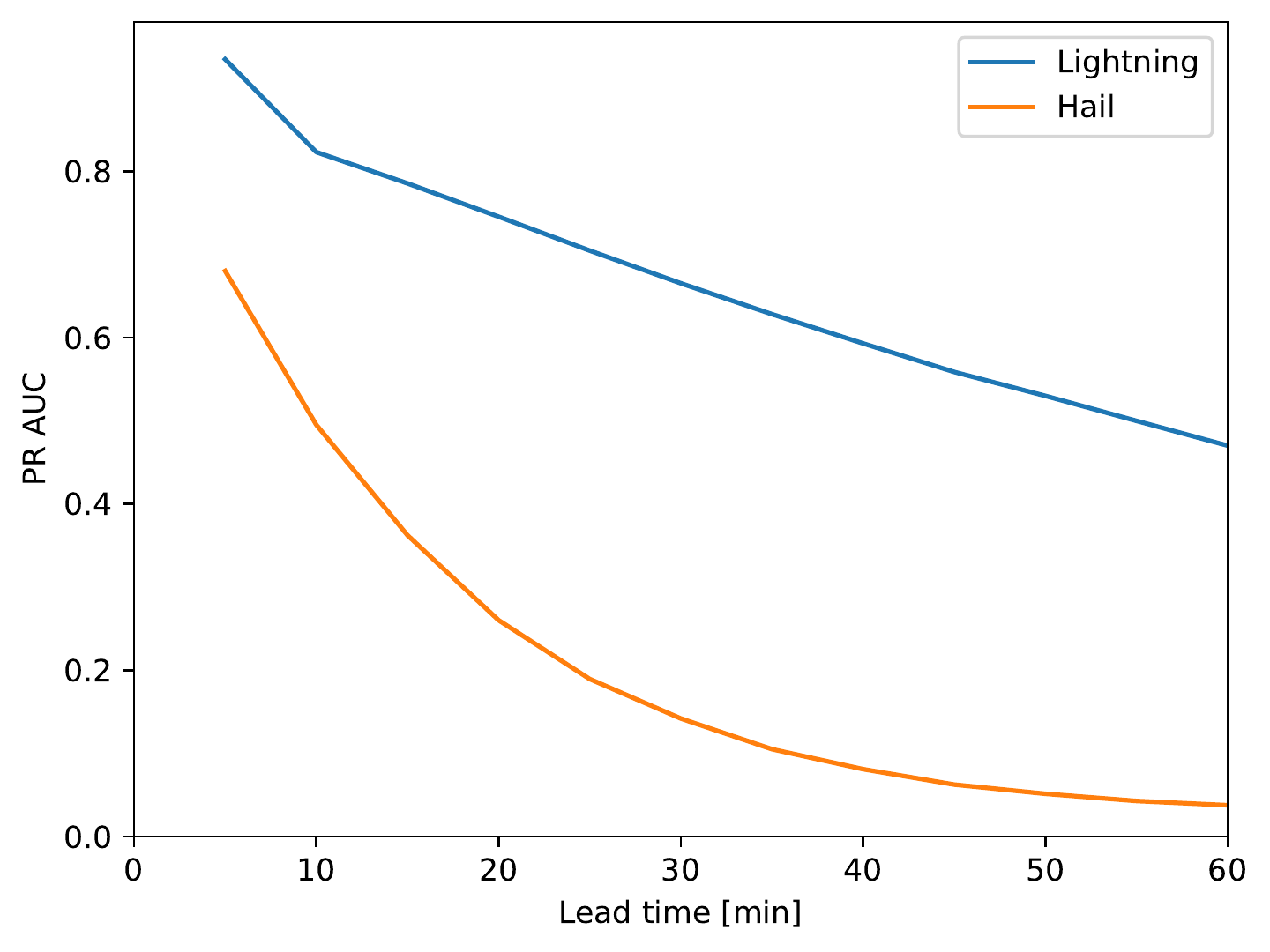}}
    \caption{As Fig.~\ref{fig:CSI-leadtime}, but for the precision--recall (PR) AUC. The AUC scores are integrated over all thresholds, so selecting a threshold is not necessary.}
    \label{fig:PR_AUC-leadtime}
\end{figure}

\begin{figure}
    \centerline{\includegraphics[width=\textwidth]{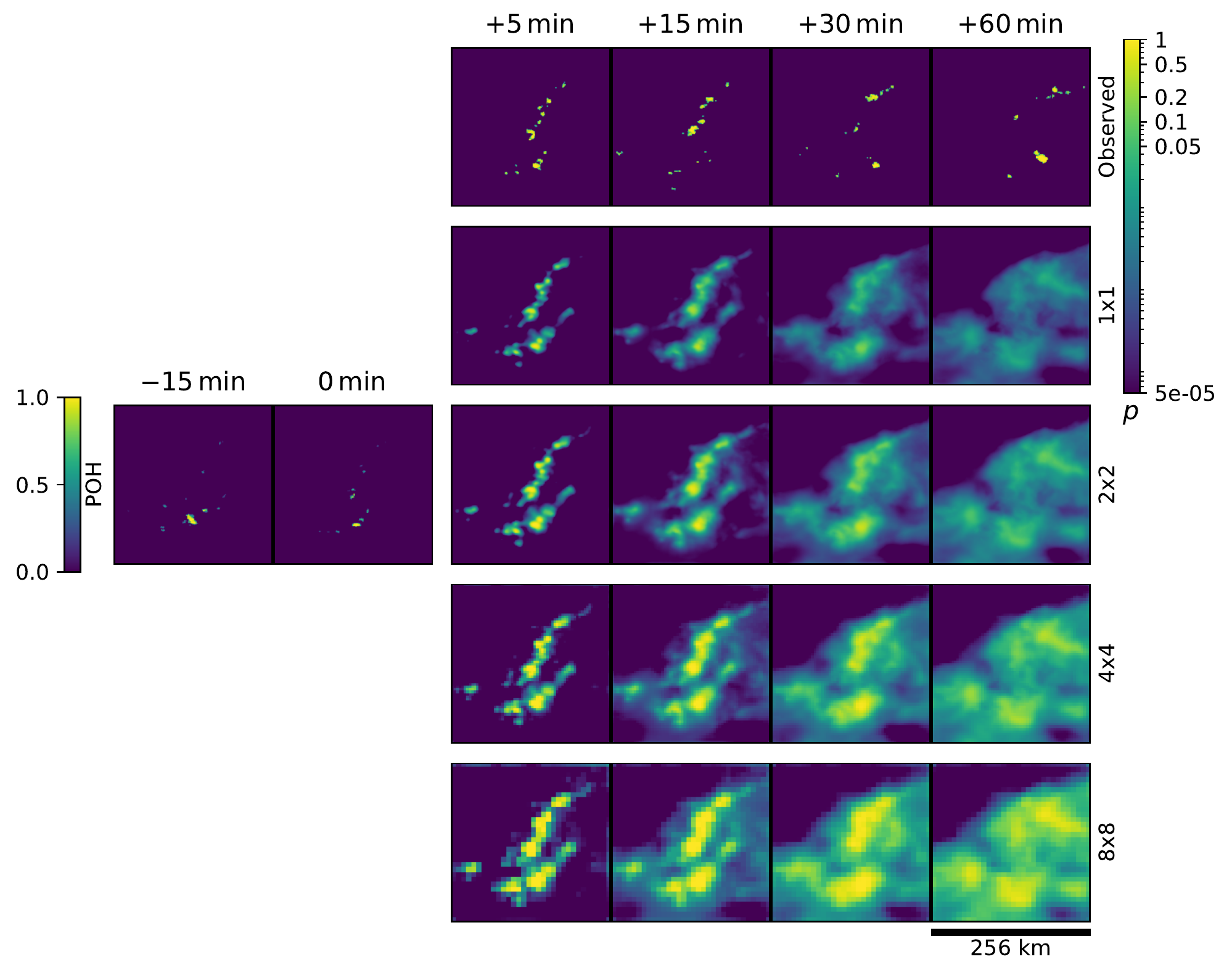}}
    \caption{The case of hail prediction from Fig. 1b of the paper at multiple scales, as indicated by the labels on the right side. ``$1 \times 1$'' is the original result, while ``$2 \times 2$'', ``$4 \times 4$'' and ``$8 \times 8$'' indicate the estimated probability that hail will occur somewhere in a square of the indicated size. The increasing probabilities at larger scales demonstrate how the model can be used to derive more confident predictions if the scale is increased. The probabilities for the larger scales have been calculated assuming that the hail probabilities between pixels are independent; in reality, they are spatially correlated so the probabilities at scales above $1 \times 1$ should be considered an estimate that is of a correct order of magnitude rather than exact.}
    \label{fig:hail-prob-scales}
\end{figure}

\noindent\textbf{Contents of this file}
%%%Remove or add items as needed%%%
\begin{enumerate}
\item Figures \ref{fig:study_area} to \ref{fig:hail-prob-scales}
\item Table \ref{table:rain-scores}
%if Tables are larger than 1 page, upload as separate excel file
\end{enumerate}
%\noindent\textbf{Additional Supporting Information (Files uploaded separately)}
%\begin{enumerate}
%\item Captions for Datasets S1 to Sx
%\item Captions for large Tables S1 to Sx (if larger than 1 page, upload as separate excel file)
%\item Captions for Movies S1 to Sx
%\item Captions for Audio S1 to Sx
%\\end{enumerate}

\noindent\textbf{Introduction}
%Type or paste your text here. The introduction gives a brief overview of the supporting information. You should include information %about as many of the following as possible (when appropriate):
% 1. a general overview of the kind of data files;
% 2. information about when and how the data were collected or created;
% 3. a general description of processing steps used;
% 4. any known imperfections or anomalies in the data.

This Supporting Information provides additional information to complement the results shown in the article ``Thunderstorm nowcasting with deep learning: a multi-hazard data fusion model''. The information provided is as follows:
\begin{itemize}
    \item Figure~\ref{fig:study_area} is a map of the study area.
    \item Figure~\ref{fig:network} illustrates the network architecture.
    \item Figure~\ref{fig:autocorr} shows the temporal autocorrelation of the lightning, hail and precipitation variables used in this study.
    \item Figures~\ref{fig:CSI-leadtime}--\ref{fig:PR_AUC-leadtime} show various metrics for the lightning and hail predictions as a function of lead time.
    \item Figure~\ref{fig:hail-prob-scales} demonstrates how predicted probabilities increase when larger scales are considered.
    \item Table~\ref{table:rain-scores} shows various metrics for the heavy precipitation predictions; these cannot be shown as a function of lead time as the prediction is made for the $60$ min accumulated precipitation.
    
\end{itemize}

\clearpage

\end{article}
\clearpage

% Copy/paste for multiples of each file type as needed.

% enter figures and tables below here: %%%%%%%
%

%
% EXAMPLE FIGURES
% ---------------
% If you get an error about an unknown bounding box, try specifying the width and height of the figure with the natwidth and natheight options.
% \begin{figure}
%\setfigurenum{S1} %%You can change number for each figure if you want, not required. "S" prepended automatically.
% \noindent\includegraphics[natwidth=800px,natheight=600px]{samplefigure.eps}
%\caption{caption}
%\label{epsfiguresample}
%\end{figure}
%
%
% Giving latex a width will help it to scale the figure properly. A simple trick is to use \textwidth. Try this if large figures run off the side of the page.
% \begin{figure}
% \noindent\includegraphics[width=\textwidth]{anothersample.png}
%\caption{caption}
%\label{pngfiguresample}
%\end{figure}
%
%
%\begin{figure}
%\noindent\includegraphics[width=\textwidth]{athirdsample.pdf}
%\caption{A pdf test figure}
%\label{pdffiguresample}
%\end{figure}
%
% PDFLatex does not seem to be able to process EPS figures. You may want to try the epstopdf package.
%

\clearpage
\begin{table}
%\settablenum{S1} %%Change number for each table
    \caption{Metrics for the prediction of heavy precipitation occurrence for $R > 10\ \mathrm{mm}$, $R > 30\ \mathrm{mm}$ and $R > 50\ \mathrm{mm}$ over the next $60$ min. The abbreviations are as in Figs.~\ref{fig:CSI-leadtime}--\ref{fig:PR_AUC-leadtime} above.} \label{table:rain-scores}
    \centering
    \begin{tabular}{l c c c}
    \hline
    Metric & $R > 10\ \mathrm{mm}$ & $R > 30\ \mathrm{mm}$ & $R > 50\ \mathrm{mm}$ \\
    \hline
    CSI & $0.320$ & $0.182$ & $0.131$ \\ 
    ETS & $0.319$ & $0.181$ & $0.131$ \\
    HSS & $0.483$ & $0.307$ & $0.232$ \\ 
    PSS & $0.909$ & $0.858$ & $0.798$ \\ 
    ROC AUC & $0.962$ & $0.930$ & $0.899$ \\ 
    PR AUC & $0.480$ & $0.252$ & $0.140$ \\
    \hline
    \end{tabular}
\end{table}

%
% ---------------
% EXAMPLE TABLE
%
%\begin{table}
%\settablenum{S1} %%Change number for each table
%\caption{Time of the Transition Between Phase 1 and Phase 2\tablenotemark{a}}
%\centering
%\begin{tabular}{l c}
%\hline
% Run  & Time (min)  \\
%\hline
%  $l1$  & 260   \\
%  $l2$  & 300   \\
%  $l3$  & 340   \\
%  $h1$  & 270   \\
%  $h2$  & 250   \\
%  $h3$  & 380   \\
%  $r1$  & 370   \\
%  $r2$  & 390   \\
%\hline
%\end{tabular}
%\tablenotetext{a}{Footnote text here.}
%\end{table}
% ---------------
%
% EXAMPLE LARGE TABLE (UPLOADED SEPARATELY)
%\begin{table}
%\settablenum{S1} %%Change number for each table
%\caption{Time of the Transition Between Phase 1 and Phase 2\tablenotemark{a}}
%\end{table}